\documentclass{article}
\usepackage{bigsky2007}
\usepackage{graphicx}
\frompage{000} \topage{000}                                              
\def\rightmark{The Heavy Ion Physics Program with ATLAS at the LHC}
\def\leftmark{N. Grau}

\title{The Heavy Ion Physics Program with ATLAS at the LHC}
\authors{
{N. Grau$^1$ {\it for the ATLAS Collaboration}
}\\[2.812mm]
{\normalsize
\hspace*{-8pt}$^1$ Columbia University, Nevis Labs, \\
Irvington, NY, 10533, USA\\[0.2ex]
}}

\abstract{The first Pb+Pb collisions at the Large Hadron Collider
(LHC) at $\sqrt{s_{NN}} = 5.52$ TeV are imminent. Heavy ion
collisions at the LHC provide an extended energy lever arm to the
existing measurements made at RHIC and SPS, especially in hard
(large-$Q^2$) processes. In this contribution an overview of the
ATLAS detector is given and the current physics focus of Heavy Ion
Working Group is discussed.}


\begin{document}

\maketitle \setcounter{page}{1}

\section{Introduction}
The Large Hadron Collider is expected to collide $p+p$ and $Pb+Pb$
collisions in the very near future at the new energy frontier for
both nuclear and high energy physics. With the many exciting results
from the hard scattering, high-$Q^2$, processes at RHIC such as
high-$p_T$ particle suppression or jet quenching\cite{PHENIXpi0} and
heavy flavor production\cite{PHENIXElectrons}, the heavy ion physics
program at the LHC is compelling. The rate for jets and heavy
flavors are substantial compared to RHIC\cite{Accardi:2002vt}. This
contribution outlines the heavy ion program of the ATLAS
detector\cite{LoI}. First, a description of the ATLAS detector is
detailed. Then a description of the physics program is outlined with
specific examples of measurements of global variables, jet
reconstruction, heavy flavor physics and low-x physics.

\section{The ATLAS Detector}
\begin{figure}[tb]
\vspace*{+2cm}
\epsfxsize=1\linewidth 
\insertplot{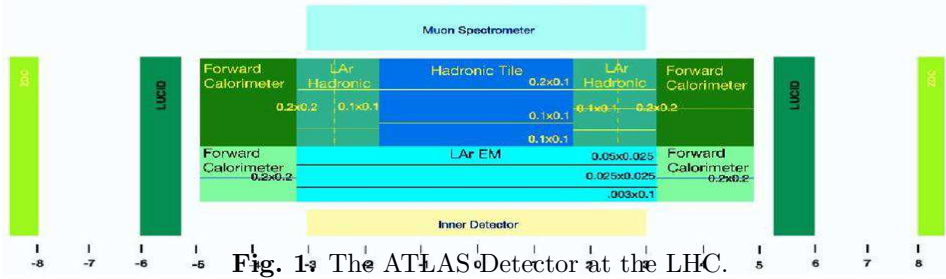} \vspace*{-5cm} \caption[]{The ATLAS Detector
at the LHC.} \label{fig1}
\end{figure}

The ATLAS detector\cite{ATLASTDR} (Fig.~\ref{fig1}) is a large,
multi-purpose detector designed for detecting and reconstructing
high-$p_{T}$ observables in $p+p$ collisions. It is composed of an
inner tracking system, electromagnetic and hadron calorimeters, and
a muon spectrometer. Even though ATLAS was designed for $p+p$
multiplicities, occupancies for central Pb+Pb HIJING events are low
enough for most subsystems for use as a heavy ion detector.

The inner tracking system covers full azimuth and approximately
$|\eta| <$ 2.5 and is within a 2T solenoidal field. It consists of
three pixel layers (r=50.5 mm, 88.5 mm and 122.5 mm), followed
outward by four layers of double-sided silicon strip detectors, and
finally the transition radiation tracker a straw tracker with 35
points for $|\eta| <$ 2.5 and $p_T
>$ 0.5 GeV. With initial studies using central HIJING events the occupancy
of the pixel layers are less than 2\% and the silicon strips are
less than 20\%. However, the occupancy for the transition radiation
tracker is large. As such, heavy ion tracking is being optimized
with the 11 space points from the pixels and the strips.

The calorimeter coverage in ATLAS is full azimuth and $|\eta| <$ 5
with several radial (longitudinal) segments. The barrel region
($|\eta| <$ 1.5) consists of a thin presampler to measure lost
energy in material before the calorimeter, three longitudinal
segments of liquid argon (LAr) electromagnetic calorimeters, and
followed by three longitudinal segments of tile hadronic
calorimeters. The endcap region (3.2 $ < |\eta| <$ 1.5 units) is
composed of the same LAr electromagnetic calorimetry as the barrel.
The hadronic calorimeter consists of two longitudinal segments of
copper plates. Finally the the forward (4.9 $ < |\eta| <$ 3.2 units)
hadronic calorimeter has three segments.

The $\eta-\phi$ segmentation of the calorimeters is dependent on the
longitudinal segment. The front (strip) layer of the barrel LAr
calorimeter is composed of cells with a typical $\Delta\eta$ x
$\Delta\phi$ of 0.003 x 0.1. Such fine granularity was built into
the calorimeter for vectoring $H\rightarrow\gamma\gamma$ events. The
middle layer of the LAr calorimeter has a typical segmentation of
0.025 x 0.025 and has the largest interaction length of the barrel.
Most electromagnetic energy is deposited within the middle layer.
The back layer of the LAr calorimeter has a typical segmentation of
0.05 x 0.025. The barrel tile calorimeter has a typical segmentation
of 0.1 x 0.1, 0.1 x 0.1, and 0.2 x 0.2 in $\Delta\eta$ x
$\Delta\phi$ in the front, middle, and back layer. In what is shown
later, towers of 0.1 x 0.1 are composed from the sum of cells from
all layers and used in jet reconstruction and elliptic flow
($v_{2}$) analyses.

Beyond the hadronic calorimeters and within a toroidal field is the
muon spectrometer, one of the largest ever constructed, with
acceptance of muons out to $|\eta| <$ 2.7 units. It consists of
three chambers containing drift tubes (in the midrapidity, barrel
region) or cathode strip chambers (in the forward rapidities) to
measure the R-z position of the tracks passing through the chambers.
The occupancy is very low since the hadronic calorimeters absorb
nearly all of the hadronic background.

The Zero Degree Calorimeter (ZDC) is located at $\eta=\pm$8 and will
consist of 2 hadronic calorimeter modules with a shower maximum
detector. These will be preceded by a highly segmented
electromagnetic calorimeter module. The ZDC is designed for
triggering and centrality definition as well as forward meson
measurements. 

\section{The Heavy Ion Physics Program$^{a}$}
With the ATLAS detector it is possible to cover a broad range of
physics observables. Currently the Heavy Ion Working Group has been
focused on global physics measurements such as $dN_{ch}/d\eta$ and
$v_{2}$, jet measurements from the longitudinally segmented
calorimeter, charm and bottom quarkonia measurements using the muon
spectrometer, and low-x physics observables utilizing the
ZDC~\cite{Steinberg:2007nm}.

\begin{figure}[b!]
\vspace*{+2cm} \epsfxsize=1\linewidth \insertplot{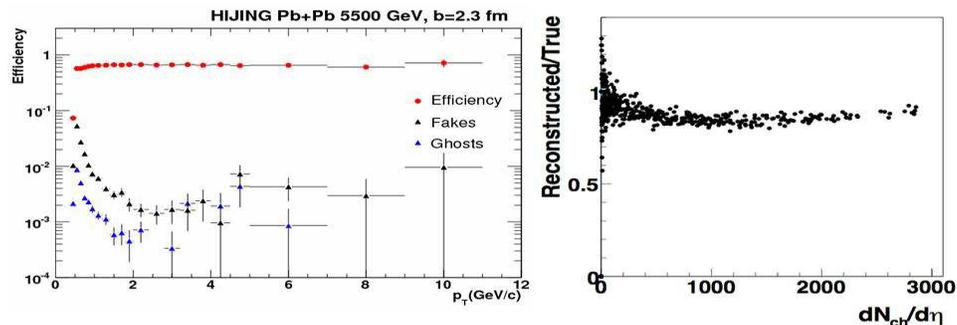}
\vspace*{-4cm} \caption[]{\textit{Left:} Efficiency, ghost rate, and
fake rates (see text) as a function of track $p_{T}$ for tracking
optimized for heavy ion events. \textit{Right:} Comparison of
reconstruction $dN_{ch}/d\eta$ from tracklets (see text) compared to
the HIJING generated $dN_{ch}/d\eta$ as a function of generated
$dN_{ch}/d\eta$.} \label{fig2}
\end{figure}

A crucial first measurement during heavy ion collisions at LHC is
the particle multiplicity density, $dN_{ch}/d\eta$. Currently, for
example, extrapolation of a logarithmic dependence of the
$dN/d\eta/\langle 0.5N_{part}\rangle$ based on measured data at
lower energies is 50\% lower than the Color Glass Condensate (CGC)
prediction\cite{CGC}. HIJING multiplicities with and without
quenching are even larger.

\begin{figure}[t!]
\vspace*{+2cm} \epsfxsize=1\linewidth \insertplot{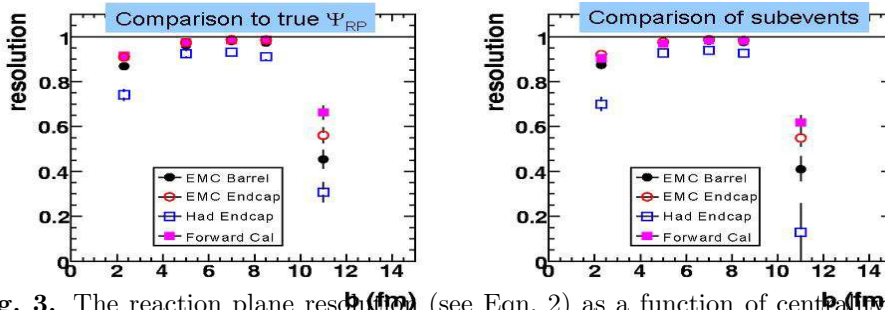}
\vspace*{-4.5cm} \caption[]{The reaction plane resolution (see
Eqn.~\ref{eq:rpres}) as a function of centrality by comparing fitted
$\Psi_{RP}$ and the generated $\Psi_{RP}$ (left) and comparing
$\Psi_{RP}$'s fitted from positive and negative $\eta$ regions
(right).} \label{fig3}
\end{figure}

Using the pixels and strips of the inner detector, it is expected
that, with good efficiency, tracking will extend below 1.0 GeV/c.
The left panel of Fig.~\ref{fig2} shows the efficiency and fraction
of ghost$^{b}$ and fake$^{c}$ tracks as a function of track $p_{T}$.
The 1.0 GeV/c cut off is artificial and considerable effort is being
made to extend the tracking to as low $p_{T}$ as possible. Early
attempts indicate that tracking down to 400 MeV/c will be possible.
Such particles pass through all layers of tracking and reach the
calorimeter with a shallow entrance angle.

An alternative approach to measure $dN_{ch}/d\eta$ including
low-$p_{T}$ particles is to perform tracklet
reconstruction~\cite{Back:2000gw} by pairing hits in the first two
pixel layers. The right panel of Fig.~\ref{fig2} shows the ratio of
measured to generated $dN_{ch}/d\eta$ as a function of generated
$dN_{ch}/d\eta$. A very good reproduction of $dN_{ch}/d\eta$ is seen
for all values of $dN_{ch}/d\eta$.

Another global observable of great interest at LHC energies is
$v_{2}$. At midrapidity, the single particle distribution can
written as
\begin{equation}\label{eq:v2}
\frac{dN}{d\phi dp_{T}} = \frac{N}{2\pi}\left(1 +
2v_{2}\cos\left(2\left(\phi-\Psi_{RP}\right)\right)\right)
\end{equation}
where $\Psi_{RP}$ is the reaction plane angle. At RHIC, large values
and scaling properties of $v_{2}$ for all measured particles
indicate that a strongly-coupled, low-viscosity partonic stage
exists at RHIC\cite{Adare:2006ti}. The non-viscous, hydrodynamical
properties of the matter created at RHIC is apparently due to the
fact that the $v_{2}/\epsilon$ is at the hydrodynamic limit. At
higher energies, a saturation of this value is expected. However,
other effects, such as path-length dependent energy loss, could
drive the observed $v_{2}$ to larger values at LHC energies. This
increasing trend of $v_2/\epsilon$ is seen in the data from SPS to
RHIC energies~\cite{hydroLimit}.

\begin{figure}[t!]
\vspace*{+2cm} \epsfxsize=1\linewidth \insertplot{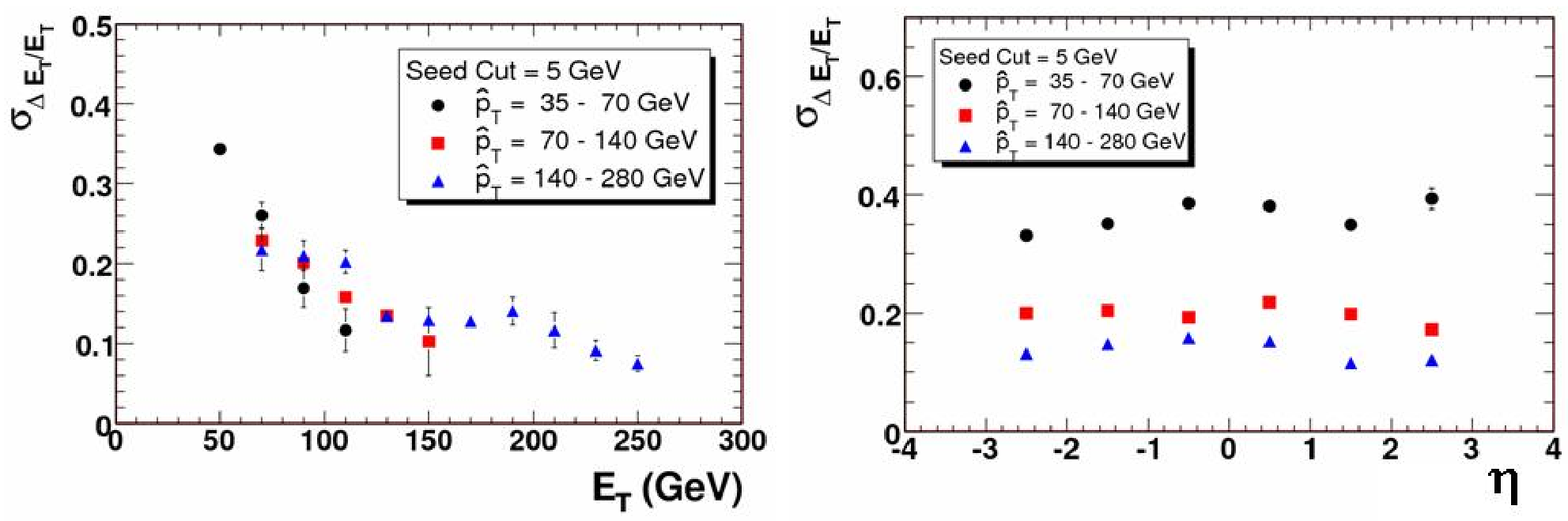}
\vspace*{-6cm} \caption[]{Energy resolution of reconstructed PYTHIA
jets embedded in b=2 fm HIJING events as a function of $E_{T}$(left)
and $\eta$(right). The PYTHIA jets were generated with the hard
scattering scale ($\hat{p}_{T}$) shown.} \label{fig4}
\end{figure}

It is possible, using the ATLAS calorimeter, to determine
$\Psi_{RP}$ and $v_{2}$ values event-by-event. A realistic flow was
simulated based on the observed RHIC data for
$v_{2}\left(p_{T},\eta,\rm{cent}\right)$. HIJING events were run
through an afterburner where all particles' azimuthal angles are
altered based on the prescription outlined in
ref.~\cite{Poskanzer:1998yz}. Calorimeter $E_{T}$ distributions are
then fit with Eqn.~\ref{eq:v2} to obtain the value of $v_{2}$ and
$\Psi_{RP}$. To evaluate the goodness of the reaction plane
determination, the resolution of the reaction plane, defined as
\begin{equation}\label{eq:rpres}
\rm{res} =
\left<\cos\left[2\left(\Psi_{RP,fit}-\Psi_{RP,gen}\right)\right]\right>
\end{equation}
is evaluated. This is shown in Fig.~\ref{fig3} for different
calorimeter regions and as a function of centrality. The right plot
of Fig.~\ref{fig3} shows the resolution based on comparison of
subevents from positive and negative $\eta$. Both resolutions are
near unity. The resolution as evaluated by the subevent technique is
consistent with the direct comparison to the generated $\Psi_{RP}$,
giving confidence that the subevent measurement of the reaction
plane is accurate.

A key strength of the ATLAS detector is the calorimetry. At LHC jets
of large energy ($E_{T} >$ 40 GeV) will be produced quite
copiously\cite{Accardi:2002vt}. With a hermetic calorimeter
event-by-event jet reconstruction is possible even within the heavy
ion underlying event. Such event-by-event information should yield
greater insights into partonic energy loss discovered at RHIC. For
complete details of the jet reconstruction and capabilities of the
ATLAS detector see ref.~\cite{WolfATLASJets}.

Jets are studied by embedding PYTHIA jets directly into HIJING
simulated events. Background subtraction schemes have been developed
to remove the large underlying event from the HIJING.  A standard
jet reconstruction technique is the cone algorithm which sums energy
within a given radius, $R = \sqrt{\Delta\phi^{2} + \Delta\eta^2}$ of
a reconstructed jet axis. Typically these algorithms are seeded by
calorimeter towers. Fig.~\ref{fig4} shows the energy resolution as a
function of jet $E_{T}$ and jet $\eta$ from a R=0.4 cone algorithm
with a seed tower of 5 GeV.

There are also two important features of the ATLAS calorimeter.
First, for the background subtraction, nearly 60\% of the underlying
event ranges out in the strip layer of the calorimeter. This is due
to the large bending of $p_{T} < $ 1 GeV/c charge particles in the
magnetic field. The bend is large enough that the particles enter
the calorimeter at a steep angle and deposit all of their energy at
the front of the calorimeter. This significantly reduces the
background in the other longitudinal calorimeter layers.

\begin{figure}[t!]
\vspace*{+2cm} \epsfxsize=1\linewidth \insertplot{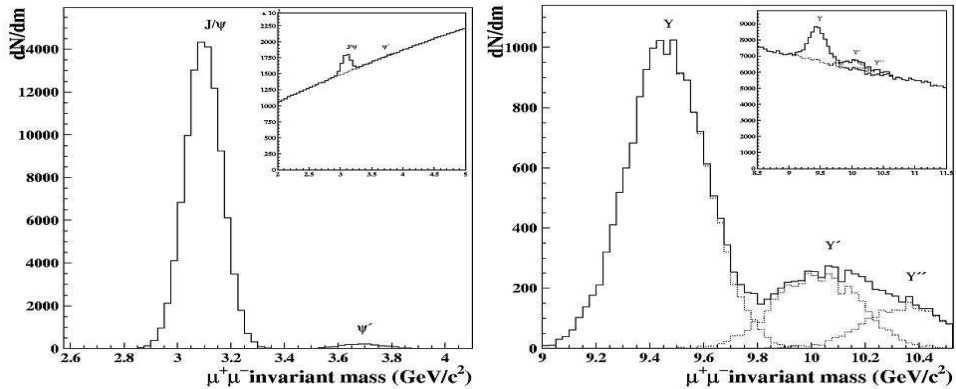}
\vspace*{-3cm} \caption[]{Di-muon invariant mass distribution for
charm quarkonia (left) and bottom quarkonia (right) expected from
the ATLAS muon spectrometer. The right plot is for $\upsilon$'s with
decay muons $|\eta| <$ 2.} \label{fig5}
\end{figure}

The other important feature of the ATLAS calorimeter is the strip
layer of the barrel electromagnetic calorimeter. The strip layer
makes it possible to detect and reject
$\pi^0\rightarrow\gamma\gamma$ decays from direct photon events.
Studies in p+p collisions with pile up indicate that a rejection
factor of 4 can be gained by information in the strip layer without
a jet isolation cut\cite{ATLASEGamma}. Initial studies from HIJING
events indicate that even in central events the occupancy in the
strip layers is so low that the background is negligible for even
fairly low photons. It is possible, then, that a rejection $>$ 1 can
be reached in heavy ion events prior to an isolation cut. Such
rejection is important since the background itself will be directly
measurable. This is an area that considerable effort is currently
being directed.

Heavy quarkonia and its suppression and regeneration is still an
unresolved topic. Currently RHIC results show a $J/\Psi$ suppression
similar to that measured at SPS\cite{Adare:2006ns}. Recent lattice
calculations indicate that different quarkonia states disassociate
at different temperatures~\cite{Asakawa:2003re}. This suggest that a
measurement of quarkonia suppression of the different states yields
the temperature of the matter produced.

The ATLAS muon spectrometer is utilized to measure and study heavy
quarkonia by their $\mu^+\mu^-$ decays. Both charm and bottom will
be accessible in ATLAS. The Di-muon invariant mass distributions are
shown in Fig.~\ref{fig5}. The resolution is sufficient to
distinguish the $J/\Psi$ and the $\Psi^{\prime}$ as well as the
$\Upsilon$ from the higher $\Upsilon$ states. Studies of the
$\Upsilon$ mass resolution and acceptance indicate the resolution is
120 MeV/c$^2$ at midrapidity, increasing at higher $\eta$. The
acceptance extends to $\pm$6 units of pseudorapidity for $\Upsilon$
down to $p_{T}$ = 0 GeV/c\cite{LaurentHeavyQuarks}.

Another avenue that is currently being explored is the photon
capability to measure the $\chi_{c}\rightarrow J/\Psi+\gamma$. This
would allow a direct measurement of the feeddown to the $J/\Psi$ and
yield an important state on the thermometer of the quark-gluon
plasma.

\begin{figure}[b!]
\vspace*{+2cm} \epsfxsize=1\linewidth \insertplot{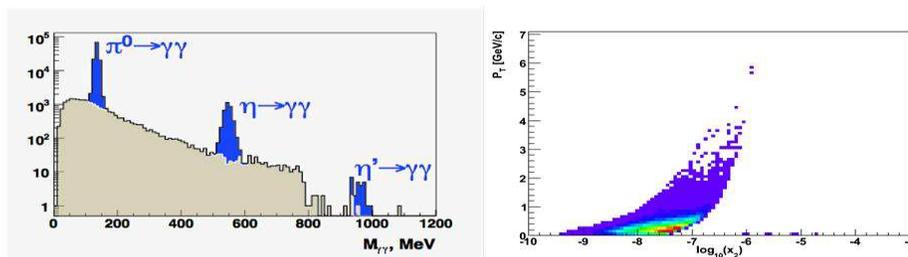}
\vspace*{-5cm} \caption[]{\textit{Left:} Di-photon invariant mass
distribution for forward produced mesons reconstructed by the ZDC.
\textit{Right:} Coverage in $p_{T}$-$x$ space of forward mesons seen
by the ZDC.} \label{fig6}
\end{figure}

A unique feature of the ATLAS ZDC is the high-resolution
electromagnetic module. The resolution is designed to detect photons
from very forward meson decays. An example of the expected di-photon
invariant mass distribution is given in the left panel of
Fig.~\ref{fig6}. The location of the ZDC at and beyond $\eta$ of
$\pm$ 8 units provides unprecedented low-$x$ coverage for A+A
collisions. A plot of the $p_{T}$ and $x$ range accessible by the
ZDC is shown in the right panel of Fig.~\ref{fig6}. At modest
$p_{T}$ values of $x$ of the order of $10^{-6}-10^{-7}$ can be
reached. This region may well be within the saturation regime making
possible study of the effects of the Color Glass Condensate.

\section{Summary}
The ATLAS detector was designed for p+p collision studies but is
well suited for studying $A+A$ collisions at the LHC. Utilizing the
unique aspects of the ATLAS detector, e.g. the calorimeter
segmentation for prompt photon measurements or the ZDC resolution
for low-x physics, a broad set of physics topics will be covered
with the ATLAS detector. This contribution has outlined a small
subset of the possible measurements with ATLAS. These measurements
may help in understanding features of RHIC data and/or they may also
uncover new and yet unforseen scenarios at the next energy frontier.

\section*{Notes}
\begin{notes}
\item[a] The figures shown here were based on studies
using modified versions of ATLAS production software. Thus, they
should be considered "ATLAS preliminary". For completeness, we list
the version of Athena software used for each figure: Fig.~2(right)
used 12.0.31 with a new tracking algorithm while Fig.~2(left) and
Figure 3 used 12.0.3. Fig.~4 used 11.0.41 and special background
subtraction algorithms.
\item[b] \textbf{Ghost} -A track which, along with multiple other tracks, matches a generated track
\item[c] \textbf{Fake} - A track with no matching generated track
\end{notes}

\vfill\eject
\end{document}